\documentclass[12pt]{article}
\usepackage{url} 
\usepackage{soul}
\usepackage{amsmath,amsfonts,amssymb}
\usepackage{authblk}
\usepackage{fullpage}
\usepackage{graphicx}
\usepackage{natbib}
\let\cite\citep

\title{Power laws in the sea ice floe size distribution: a stochastic theory}

\author[1,2]{Samuel N. Stechmann\thanks{Corresponding author, stechmann@wisc.edu}} 
\author[1]{Jiuhua Hu\thanks{Current address, Virginia Polytechnic Institute and State University, Blacksburg, Virginia, USA}}
\author[3]{Brandon P. Montemuro\thanks{Current address, Saint Michael's College, Colchester, Vermont, USA}}
\author[1]{Nan Chen}
\author[3]{Georgy E. Manucharyan}
\author[1]{Evelyn Tollar}
\author[1]{Yujia Zhang}

\affil[1]{Department of Mathematics, University of Wisconsin--Madison, Madison, Wisconsin, USA}
\affil[2]{Department of Atmospheric and Oceanic Sciences, University of Wisconsin--Madison, Madison, Wisconsin, USA}
\affil[3]{School of Oceanography, University of Washington, Seattle, Washington, USA}

\date{March 30, 2026}

\begin{document}

\maketitle

\begin{abstract}
Sea ice is a complex system, and
observations have shown that ice segments (i.e., floes) have
a wide range of sizes, with a floe size distribution
that follows a power law.
However, a theory for the power law and its exponent
have remained elusive.
Here, floe-resolving numerical simulations
are investigated with a discrete element model,
in order to gain further information by gathering statistics of fracture and welding events.
Then, based on the insights from the floe-resolving simulations,
a stochastic fragmentation--coagulation theory
is proposed.
Exact solutions are found with a power law. 
The power-law exponent can take a variety of values, and it depends on the fracture and welding rates.
Such behavior is reminiscent of seasonal changes in the power-law exponent,
which have been reported in past analyses of observational data.
\end{abstract}



%
%

\section{Introduction}
\label{sec:intro}

Sea ice is an important component of the climate system \cite{barry1993arctic,vihma2014effects,bacon2023arctic,kusahara2025causes},
with implications for
ecology \cite{post2013ecological,arrigo2014sea,meier2014arctic},
transportation \cite{melia2016sea,pizzolato2016influence,saenko2025large}, 
and biogeochemistry \cite{thomas2003biogeochemistry,lannuzel2020future}.

Sea ice provides striking examples of multiscale features \cite{golden2020modeling,webster2022observing,deng2024particle}, and it is composed of individual pieces of ice known as floes, which can range in size from meters to tens of kilometers. Floe-scale interactions cumulatively affect macro-scale sea ice dynamics, commonly represented in climate models as a continuous material with idealized viscous-plastic rheology \cite{hibler1979dynamic}. However, continuous rheologies do not explicitly simulate important granular aspects of sea ice, like the floe size distribution \cite{rothrock1984measuring} and the associated jamming behavior \cite{beltaos2009field,behringer2018physics,damsgaard2018application,montemuro2025role} that can lead to intermittency and threshold behavior in macro-scale transport. As an alternative to continuous sea ice modeling, which is computationally efficient enough to make long-term global climate projections, discrete element modeling (DEM) \cite{cundall1979discrete,potyondy2004bonded} provides a more detailed view into the floe-scale dynamics by explicitly simulating collision physics and providing a framework for fracture and welding processes that can naturally result in equilibrated macroscale sea ice characteristics \cite{hopkins2004formation,wilchinsky2010effect,herman2013numerical,herman2016discrete,Kulchitsky2017siku,damsgaard2018application,liu2018ice,tuhkuri2018review,west2021bonded,manucharyan2022subzero,montemuro2025zenodo}. 

A key characteristic of interest is the floe size distribution (FSD). Subject to external forcing from atmospheric winds and oceanic currents, floes actively form during freezing, weld together to create consolidated clusters of floes, and fracture upon collisions. These processes create a variety of floe sizes. The FSD has been analyzed in a wide range of observational studies, segmenting floes from satellite sea ice observations in different geographic locations and seasons. These observations indicate that the FSD approximately obeys a power-law distribution with exponents ranging from about -3 to -1 \cite{rothrock1984measuring,holt2001effect,horvat2019estimating,denton2022characterizing}. Furthermore, recent studies demonstrated that DEMs could emulate the observed power-law formation in simulations where floes undergo fracturing and welding processes representing winter sea ice or only fracturing caused by collisions representing summer dynamics \cite{manucharyan2022subzero,montemuro2025role}. 

\begin{figure}[htb]
    \centering
    \includegraphics[width=0.65\textwidth]{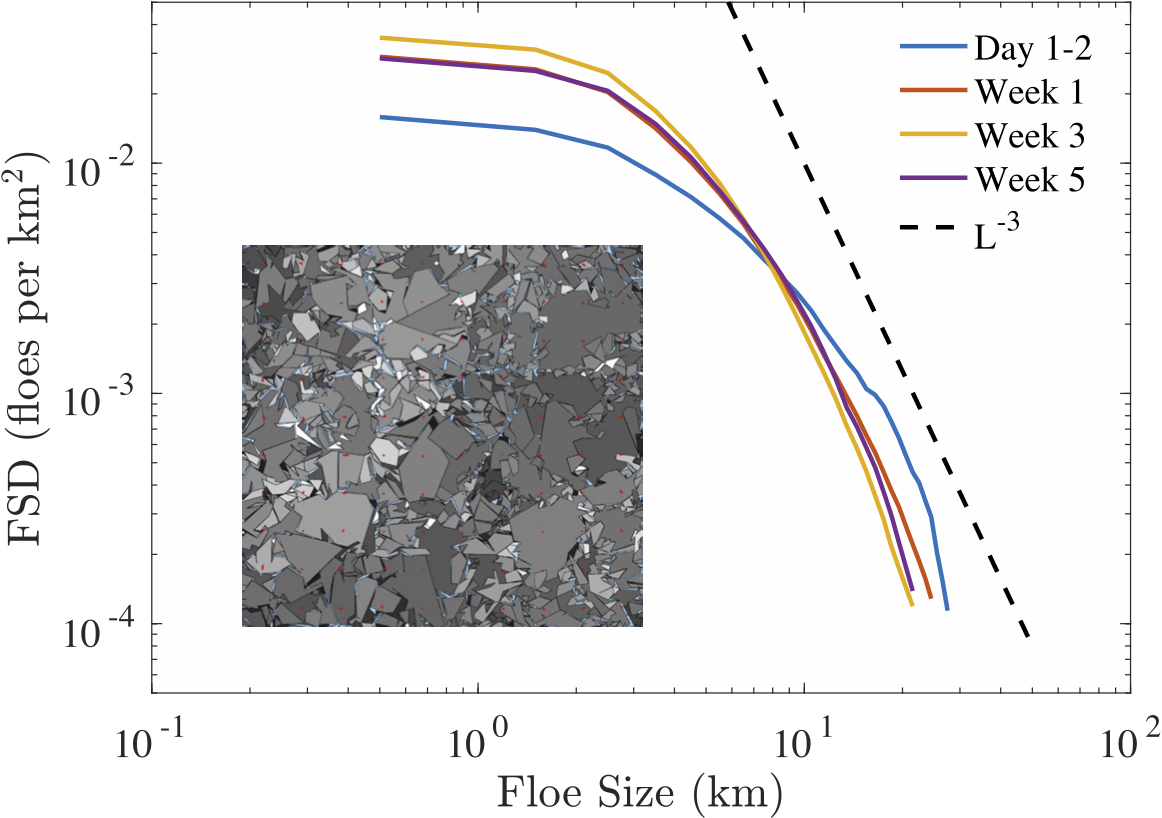}
    \caption{Equilibration of the FSD toward a power law in a SubZero DEM simulation of winter sea ice, where fracturing, welding/ridging, and formation of ice floes are active \cite{manucharyan2022subzero}. The power-law exponent of -3 is plotted for reference. Inset represents the model domain of 100 km in size with floes colored by their thicknesses, ranging from newly formed sea ice (black) to about 2 m (white). Adapted from \citet{manucharyan2022subzero}.}
    \label{fig:floe-snapshot-and-fsd}
\end{figure}

To gain a better understanding of the FSD, a variety of different mechanisms have been investigated for their influence on the FSD \cite{zhang2015sea,horvat2017evolution}, and some exactly solvable theories have been proposed for the FSD. An early, interesting idea was to draw on an analogy with rock fractures \cite{weiss2001fracture} and to model ice fractures using a fractal construction procedure \cite{toyota2011size}; it could produce ice floe populations with power laws, although the exponent values are constrained and do not represent the full range of exponents seen in observational data. Other exactly solvable models predict an FSD that follows a truncated Pareto or inverse gamma distribution \cite{herman2010sea} and a log-normal distribution \cite{montiel2022theoretical} rather than a power law. These studies also present observational evidence that the FSD may follow an inverse gamma or log-normal distribution in some circumstances \cite{herman2010sea,montiel2022theoretical}, while the power-law paradigm is in widespread use in many scenarios \cite{rothrock1984measuring,holt2001effect,horvat2019estimating,denton2022characterizing}.

Despite these advances, a basic theory explaining the emergence of power laws from fracture and welding processes and predicting the power-law exponent is currently missing. Furthermore, a desirable property of such a theory is that the power-law exponent can have a wide range of values as seen in observational data, which is a particular feature of interest for sea ice.

Here, we investigate simple models for fracturing and welding of sea ice floes to develop a basic theory for power laws in the FSD. 

Inspiration for simple models is also provided by other settings,
where concepts from stochastic modeling and statistical physics 
have been valuable for understanding a variety of weather and climate processes.
For instance,
clouds and rainfall can be viewed from the perspectives of critical phenomena and phase transitions 
\cite{peters2006critical,peters2010universality,stechmann2014first,hottovy2015spatiotemporal,neelin2017global},
and sea ice properties have been linked with concepts such as percolation theory and the Fokker--Planck equation \cite{golden1998percolation,toppaladoddi2015theory,popovic2018simple}.

In section~\ref{sec:dem} below, we first investigate floe process statistics in floe-resolving numerical simulations. Informed by these results, a stochastic theory for the FSD power law is proposed in section~\ref{sec:theory}. Conclusions are summarized in section~\ref{sec:conclusions}.

\section{Floe-resolving numerical simulations for fracture and welding event statistics}
\label{sec:dem}

\begin{figure}[htb]
    \centering
    \includegraphics[width=0.75\textwidth, clip, trim={2cm 0 1.5cm 0}]{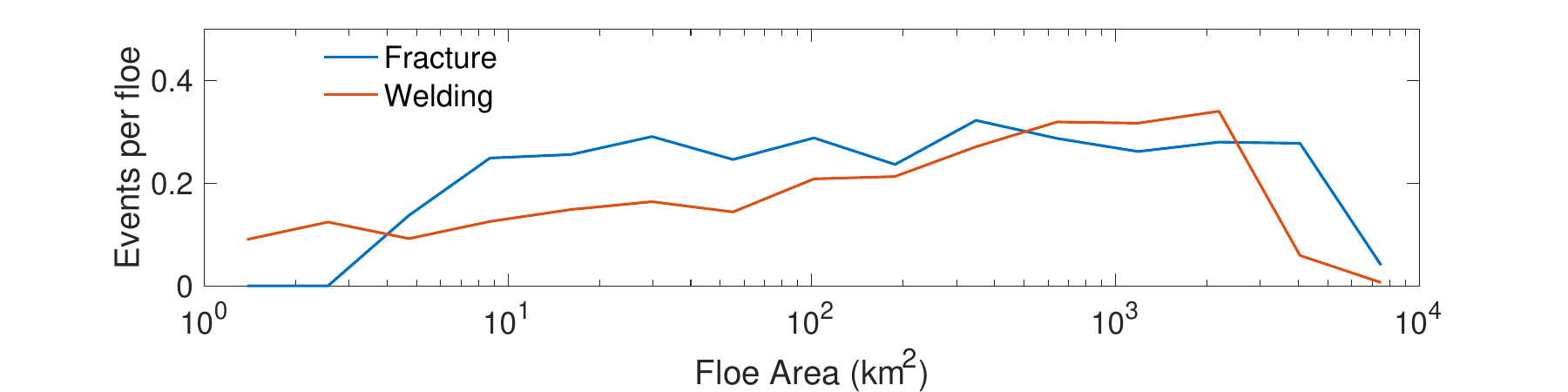}
    \caption{Fracture and welding rates plotted as the average number of events per floe. The floes are grouped by logarithmically-spaced area bins, computed for the idealized SubZero simulation where only fracturing and welding processes are present (see Supporting Information). The DEM simulation suggests that the fracture and welding rates per floe have little or no dependence on floe size.}
    \label{fig:rates-per-floe}
\end{figure}

To help inform the theoretical model setup, we first use a sea ice DEM, SubZero \cite{manucharyan2022subzero,montemuro2025zenodo}, in an idealized configuration where floe fractures and welding occur in a confined domain subject to prescribed external forcing by uniform winds and freezing atmospheric temperatures (see Supporting Information). This idealized DEM setup allows us to gain insights into the fundamental statistical nature of fracture and welding events and justify some assumptions in our analytical model. Consistent with more complex prior model simulations of winter-like sea ice (see Fig. \ref{fig:floe-snapshot-and-fsd}), the power law-like FSD emerges in our setup after the initially homogeneous-in-size floes have undergone multiple fractures and welding events. 
Statistically, the events that affect the FSD can be characterized by fracture and welding rates, which are probabilities that floes of a given size fracture or weld. While these rates can, in principle, be size-dependent, the DEM simulation demonstrates that this dependence is not overly strong (Fig. \ref{fig:rates-per-floe}). To develop a basic theory, we consider those rates to be scale-independent.

\section{A stochastic theory for power laws in the floe-size distribution}
\label{sec:theory}

Here, we present an idealized stochastic model that considers the fracturing and welding of sea ice floes to explain the emergence of a power-law in the FSD. Based on the insights from the DEM simulation,
we formulate a type of fragmentation--coagulation model.
Two constant parameters will specify the model:
\begin{subequations}
\begin{align}
    r_f &= \mbox{fracture rate, per floe}, \\
    r_w &= \mbox{welding rate, per floe},
\end{align}
\label{eqn:rf-rw}
\end{subequations}which are constants and independent of floe size,
as indicated by the DEM simulation statistics in
Figure~\ref{fig:rates-per-floe}. In analogy with chemical reactions, 
fracture and welding events can be described symbolically as
\begin{align}
    S_j &\rightarrow 2S_{j+1}, 
    \label{eqn:chem-frac}
    \\
    S_j &\leftarrow 2S_{j+1},
    \label{eqn:chem-weld}
\end{align}
respectively, which indicate that a fracture event will
transform a floe of size category $j$ into two floes of
size category $j+1$, and vice versa for a welding event.
For simplicity we assume that a fracture event will
break a floe into two equal-sized pieces, and that a welding
event will weld together two equal-sized floes.
(Some generalizations are possible; see, for instance,
\eqref{eqn:power-law-sol-c} below.)
The floe area $A_j$ for each size category $j$ 
can then be described as
\begin{equation}
    A_j = A_* 2^{-j}, \quad j=0,1,2,\cdots, J,
    \label{eqn:Aj-def}
\end{equation}
where $A_*$ is a maximum floe area and $A_* 2^{-J}$
is a minimum floe area. The state of the collection of
ice floes is described by
\begin{equation}
    \eta(j,t) = \mbox{number of floes of size category}\; j,
    \label{eqn:eta-def}
\end{equation}
which is a stochastic quantity. The evolution of $\eta(j,t)$
is influenced by randomly occurring fracture and welding events.
At each time step, the transition probabilities for fracture and welding events are
\begin{align}
    &\mathbb{P}\{ S_j\rightarrow 2S_{j+1} \;\mbox{in}\; (t,t+\Delta t] \} = r_f \,\eta(j,t)\, \Delta t, 
    \nonumber \\
    &\qquad \mbox{for all} \; j=0,1,\cdots, J-1, \;\mbox{and}
    \label{eqn:trans-prob-frac} \\
    &\mathbb{P}\{ 2S_j\rightarrow S_{j-1} \;\mbox{in}\; (t,t+\Delta t] \} = r_w \, \eta(j,t) \, \Delta t, 
    \nonumber \\
    &\qquad \mbox{for all} \; j=1,2,\cdots, J.
    \label{eqn:trans-prob-weld}
\end{align}  
where $\Delta t$ is a small time step and $t$ is the time.
These transition probabilities are in accordance with 
\eqref{eqn:rf-rw}: if the fracture rate per floe is $r_f$,
then the fracture rate for the population of 
all size-category-$j$ floes is $r_f \,\eta(j,t)$, 
and similarly for welding
(see further notes on this below \eqref{eqn:dfjdt}).
In terms of $\eta(j,t)$ evolution, a fracture event 
brings a transition of
\begin{align}
    \eta(j,t+\Delta t) &= \eta(j,t)-1,
    \nonumber \\
    \eta(j+1,t+\Delta t) &= \eta(j+1,t)+2,
    \nonumber \\
    & \mbox{for} \; S_j\rightarrow 2S_{j+1},
\end{align}
and a welding event brings a transition of
\begin{align}
    \eta(j,t+\Delta t) &= \eta(j,t)-2,
    \nonumber \\
    \eta(j-1,t+\Delta t) &= \eta(j-1,t)+1,
    \nonumber \\
    & \mbox{for} \; 2S_j\rightarrow S_{j-1}.
    \label{eqn:weld-eta-adjust}
\end{align}

From this stochastic process, one can also define its expected value,
\begin{equation}
    f_j(t) = \mathbb{E}[\eta(j,t)].
\end{equation}
The evolution equation for the expected value, $f_j(t)$,
can be derived in two ways 
(see Supporting Information for details):
from a random-time-change
representation of the stochastic jump process \cite{anderson2011continuous},
or from the master equation, i.e., the Kolmogorov forward equation
\cite{gardiner2021elements}.
The resulting evolution equation for $f_j(t)$ is
\begin{equation}
    \frac{df_j}{dt}
    =2r_f f_{j-1}-(r_f+2r_w)f_j+r_w f_{j+1}.
    \label{eqn:dfjdt}
\end{equation}
Note that this equation is linear, in accordance with the linear rates in \eqref{eqn:trans-prob-frac}--\eqref{eqn:trans-prob-weld}, whereas in many applications of coagulation (or coalescence, aggregation, or chemical reactions) the evolution equation is quadratic \cite{ball1990discrete,wattis2006introduction,anderson2011continuous}. A quadratic equation often arises because the coagulation rate is commonly determined by the probability of (relatively rare) collisions of two particles. In contrast, for sea ice, the floes are regularly in contact with other floes, and it is reasonable to model the welding rate as done herein as being determined by other factors, such as environmental factors (such as temperature).

The model in \eqref{eqn:dfjdt} is exactly solvable.
The steady state scenario, 
\begin{equation}
    0
    =2(r_f f_{j-1}-r_w f_j)-(r_f f_j-r_w f_{j+1}) \, ,
    \label{eqn:steady-f}
\end{equation}
is a valuable case.
A solution can be found, in the spirit of detailed balance conditions \cite{gardiner2021elements},
by requiring each of the two terms in \eqref{eqn:steady-f}
to vanish individually.
One can see that
\begin{equation}
    f_{j+1}=\frac{r_f}{r_w}f_j
\end{equation}
is a solution, which, upon recursive multiplication,
can be written as
\begin{equation}
    f_j=\left(\frac{r_f}{r_w}\right)^j f_0 \, .
    \label{eqn:fj-sol}
\end{equation}
Recall now, from \eqref{eqn:Aj-def}, that size category $j$
corresponds to a floe area of $A_j=2^{-j}$.
Hence, rewriting \eqref{eqn:fj-sol} in terms of $2^{-j}$
yields 
\begin{equation}
    f_j=2^{j\log_2 \left(\frac{r_f}{r_w}\right)} f_0 \, ,
\end{equation}
or, in terms of area,
\begin{equation}
    f(A)=A^{-\alpha} f(A_*),
    \quad\mbox{where}\quad 
    \alpha=\log_2 \frac{r_f}{r_w} \, ,
    \label{eqn:power-law-sol}
\end{equation}
which is a power-law solution for the FSD.

More generally, one can begin at 
\eqref{eqn:chem-frac}--\eqref{eqn:chem-weld} 
with $c$ floes welding into one floe and, in reverse,
one floe fracturing into $c$ floes. In this case,
following the same derivation, one finds
\begin{equation}
    f(A)=A^{-\alpha} f(A_*),
    \;\mbox{where}\; 
    \alpha=\log_c \frac{r_f}{r_w} \, ,
    \;\mbox{and}\; A_j=c^{-j} \, ,
    \label{eqn:power-law-sol-c}
\end{equation}
where the base of the logarithm is $c$ in this more general case.

\begin{figure}[htb]
\centering
\includegraphics[width=0.75\textwidth,  clip, trim={2cm 6.5cm 1.5cm 7.2cm}]{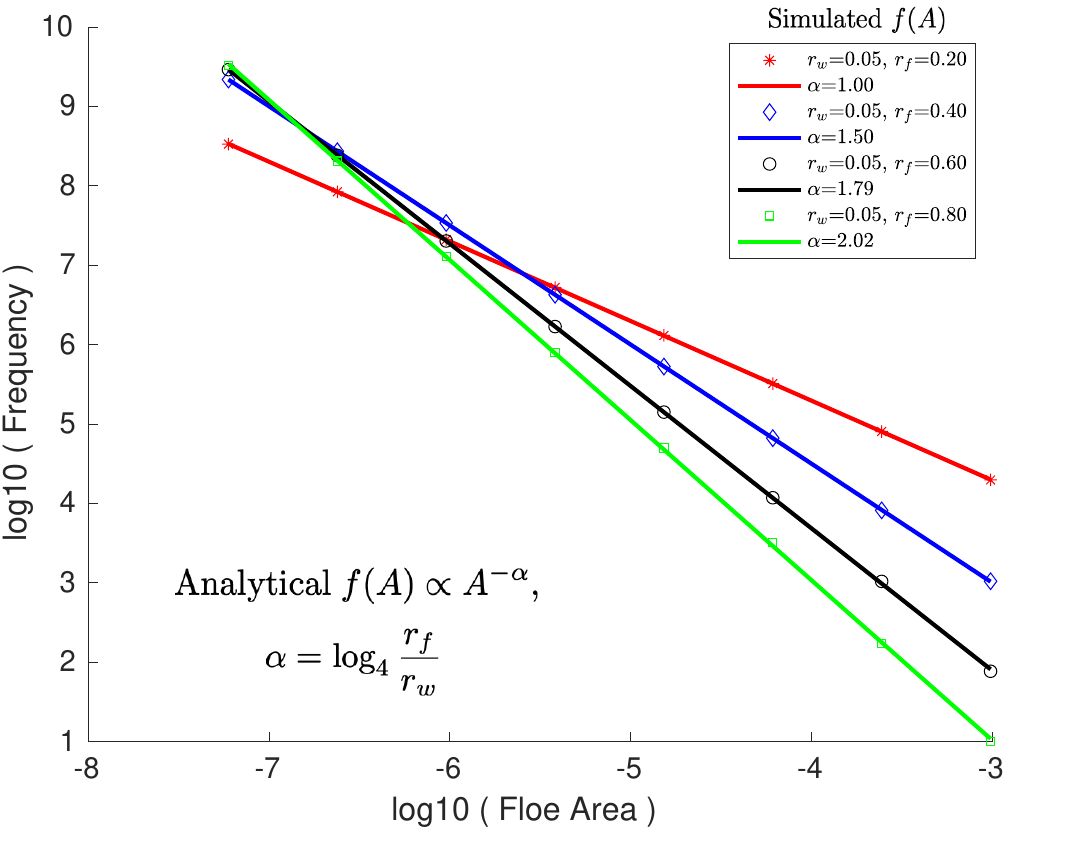}
\caption{Log--log plots of the time-averaged FSD, from numerical simulations of the stochastic model in \eqref{eqn:eta-def}--\eqref{eqn:weld-eta-adjust}.
Each simulation involves a large number of floes: approximately $10^{9}$.
The lines show the best-fit of a power law for the simulated data,
and the best-fit value of the power-law exponent $\alpha$ is shown in the legend.
An inset shows the analytical formulas for $f(A)$ and $\alpha$ 
from \eqref{eqn:power-law-sol} and \eqref{eqn:power-law-sol-c} with $c=4$.
The plots illustrate the power law, and also illustrate that the power-law exponent can take any value, depending on the fracture and welding rate parameters, $r_f$ and $r_w$. }
\label{fig:simple-sim-power-law}
\end{figure}

The power-law exponent $\alpha$ in \eqref{eqn:power-law-sol}
is seen to depend on $r_f$ and $r_w$, the rates of
fracture and welding, respectively.
An illustration is shown in Figure~\ref{fig:simple-sim-power-law}. 
Numerical simulations of the stochastic model were conducted for various values of
$r_f$ and $r_w$ (see Supporting Information for further description), 
and the simulated FSD is in agreement with the theory.
For a larger fracture rate $r_f$, the exponent $\alpha$ is larger
and the FSD has a more rapid decay. In even simpler terms,
as physical intuition would suggest, more fractures will create
more small floes and a rapidly decaying FSD.

In observational data, a variety of different power-law exponents
has been seen, depending on various factors.
In a recent study of a relatively comprehensive dataset \cite{denton2022characterizing},
different power laws were seen, and
the power-law exponent $\alpha$ was in the range of
1.65 to 2.03. This range of $\alpha$ values is 
similar to the cases of the stochastic model that are illustrated in
Figure~\ref{fig:simple-sim-power-law}.
The stochastic model here has the desirable property
that it can have a power law with any value
of the exponent $\alpha$, depending on the rates of
fracture and welding.

In nature, 
the FSD exponent $\alpha$ can increase as the
seasons transition from spring and through summer,
as fractures, welding, and other processes also undergo
seasonal variations
\cite{perovich2014seasonal,hwang2017winter,stern2018seasonal,stern2018reconciling,denton2022characterizing,buckley2024seasonal,moncada2025comparing}.
In viewing these effects in the context of the simple model
here, the state of the FSD can be in a statistical 
equilibrium on short time scales, and in the form of
the power-law formula in \eqref{eqn:power-law-sol};
and the fracture and welding rate parameters $r_f$ and $r_w$
can be slowly evolving parameters that evolve on a 
seasonal time scale, leading to a power-law exponent $\alpha$
that also evolves on the seasonal time scale.

As an additional experiment, we have investigated
how the FSD might look in the SubZero DEM
if the DEM were forced to have 
fracture and welding rates that are exactly constant.
This is in contrast to the approximately constant rates that emerged
from the SubZero DEM simulations in Figure~\ref{fig:rates-per-floe}.
Since the rates are emergent properties of a realistic SubZero DEM simulation,
it is difficult to probe relationships between rates and FSD power laws
in realistic SubZero DEM simulations.
In the scenario of exactly constant rates, it is straightforward to probe
the relationship between the rates and the FSD power laws,
in a scenario with some of the complexity of SubZero DEM simulations,
such as irregular floes of general shapes.
Additional aspects of the setup and the results are described in the Supporting Information.
The results of the simulations show that a power law emerges,
and the power law exponent increases as the rate ratio $r_f/r_w$ increases.
The simulated relationship between the rates and the FSD power law
is broadly similar to the theoretical prediction in \eqref{eqn:power-law-sol}
with some deviations due to the additional complexity such as
irregular floe shapes of the SubZero DEM model.
In sum, these additional experiments
provide further evidence that the simple mechanisms of the 
exactly solvable model here may be at work in the more complex
setting in nature.

\section{Conclusions}
\label{sec:conclusions}

In summary, a stochastic model was investigated for
the fracture and welding of sea ice floes,
and the model has exact solutions for the FSD in the form of power laws.
The power-law exponent in the model can take any value,
which is consistent with the appearance of different power-law
exponents in analyses of different observational datasets,
and consistent with
the concept of seasonal changes in the power-law exponent.
The model's power-law exponent depends on two parameters:
the rates of fracture and welding.
The setup of the stochastic model was informed by DEM simulations
which provide detailed information of floe fractures
and welding.
The fracture and welding are analogous to 
a type of fragmentation--coagulation process.
The stochastic model framework could be useful for 
interpreting observational data and 
observed power-law properties, and it provides
further motivation for observational studies of 
fracture and welding processes and other
floe-scale processes.

\section*{Software}
The software used to perform calculations have been made available at\\
https://doi.org/10.5281/zenodo.7199940 \cite{montemuro2025zenodo}
and are currently also being provided at\\ https://github.com/SeaIce-Math/SubZero
and https://github.com/Caltech-OCTO/Subzero.jl

\section*{Acknowledgments}

This research is partially supported by
grant ONR MURI N00014-19-1-2421.

This material is based upon work supported by the
National Science Foundation under Grant No. DMS-2424139,
while author S.N.S. was in residence at the
Simons Laufer Mathematical Sciences Institute
in Berkeley, California, during the Fall 2025 semester.

The authors thank Shukai Du, Changhong Mou and Melisa Erman
for discussions of the models in the early stages of the project.

The authors thank the
Wisconsin Science and Computing 
Emerging Research Stars (WISCERS) program
for encouraging and organizing the participation of
undergraduate researchers
Melisa Erman, Evelyn Tollar, and Yujia (Marissa) Zhang.

%
%

\bibliographystyle{abbrvnat}
\bibliography{agureferences}

%
%
%
%
%

\clearpage
\appendix
\section*{Supporting Information}







\section{Simulation setup: Subzero model}

In the main text, the SubZero discrete element model (DEM) was used in an idealized configuration to allow us to gain insight into the fundamental statistical nature of fracture and welding events and justify some assumptions in our analytical model. The setup of the simulation is as follows.

The DEM was initiated by placing 250 floes in a fully packed domain measuring 400 km in length and 200 km in width. These floes have a uniform thickness of 0.25 m. The wind in the simulation blows at a speed of 13 m/s from west to east. We set the oceanic stresses to zero for this simplified test. The floes are subjected to the fracture and welding processes without corner grinding, floe shape simplification, ridging, or rafting. Floe fractures and welding follow the processes described in \cite{manucharyan2022subzero}, where fractures specifically follow the Hibler ellipse method described. The parameter values for welding, fracturing, and other values used in this paper that may deviate from \cite{manucharyan2022subzero} are provided in Table \ref{tbl:Subzero_parameters}. As the eastward wind causes floes to exit the domain at its eastern side, new floes are only created when the western ice edge moves 50 km to maintain a relatively constant stress in the simulation. We have modified the SubZero code to monitor individual fracturing and welding events. The floes are categorized based on their size, and we record each occurrence of an event (fractures and welding) for a floe within that bin throughout the simulation.

\begin{table}[ht]
\caption{ A list of key parameters used in the SubZero model, including their default numerical values, a brief description, and the processes that use these parameters.}
\footnotesize
\centering

\vspace{10pt}
\begin{tabular}{ p{3cm} p{5cm} p{3cm}  }
Parameter   & Description & Process \\
\hline
 $E$ = 3.5 $\times$ 10$^7$ Pa   & Young's Modulus & Floe Interactions \\
$G = \frac{E}{2(1+\nu)}$ &   Shear Modulus & \\
$\nu = 0.3$ &   Poisson's ratio & \\
 $\mu = 0.2$ & Coefficient of Friction & \\
 \hline
 $N_{Frac}$=100  & Time steps between fracturing   & Physical Processes\\
 $N_{Pieces}=3 $ & Number of pieces for fracturing & \\
 $P^*$ = 1.25$\times$10$^5$  N m$^{-1}$    & Floe strength-to-thickness ratio& \\
  $N_{Frac}$=25  & Time steps between welding   & \\
 \hline
 $\Delta t$ =10 s & Integration time step & Time stepping\\

  \hline
\end{tabular}
  \label{tbl:Subzero_parameters}
\end{table}

\section{Derivations of the evolution equation for the expected value of floe size distribution: Stochastic sea ice floe model}

In this section, two derivations are shown for the 
evolution equation for $f_j(t)$, i.e., for the expected value of the number of floes in size category $j$,
in the stochastic sea ice floe model. 
One derivation begins
from a random-time-change
representation of the stochastic jump process for $\eta(j,t)$ \cite{anderson2011continuous},
and the other derivation begins from the master equation, i.e., the Kolmogorov forward equation
\cite{gardiner2021elements}.

\subsection{Derivation \#1}

The starting point is the evolution equation
for the stochastic state $\eta(j,t)$, or $\eta_j(t)$.
Recall that $\eta(j,t)$ is the number of floes in size category $j$ at time $t$,
and it evolves in time due to stochastic events of fracture and welding.
Its stochastic evolution can be described in the random-time-change representation as
\begin{align}
    \left(
    \begin{array}{c}
    \eta(0,t) \\ \eta(1,t) \\ \eta(2,t) \\ \vdots \\ \eta(J,t)
    \end{array}
    \right)
    &=
     \left(
    \begin{array}{c}
    \eta(0,0) \\ \eta(1,0) \\ \eta(2,0) \\ \vdots \\ \eta(J,0)
    \end{array}
    \right)
    \nonumber \\
    &\quad +
    Y_{f,0}(\int_0^t R_{f,0}(\eta_0(t')) \, dt')
    \left(
    \begin{array}{c}
    -1 \\ 2 \\ 0 \\ \vdots \\ 0
    \end{array}
    \right)
    +
    \cdots
    +
    Y_{f,J-1}(\int_0^t R_{f,J-1}(\eta_{J-1}(t')) \, dt')
    \left(
    \begin{array}{c}
    0 \\ \vdots \\ 0 \\ -1 \\ 2
    \end{array}
    \right)
    \nonumber \\
    &\quad +
    Y_{w,1}(\int_0^t R_{w,1}(\eta_1(t')) \, dt')
    \left(
    \begin{array}{c}
    1 \\ -2 \\ 0 \\ \vdots \\ 0
    \end{array}
    \right)
    +
    \cdots
    +
    Y_{w,J}(\int_0^t R_{w,J}(\eta_{J}(t')) \, dt')
    \left(
    \begin{array}{c}
    0 \\ \vdots \\ 0 \\ 1 \\ -2
    \end{array}
    \right),
    \label{eqn:eta-vec-evol}
\end{align}
where $Y_{f,0}, \cdots, Y_{f,J-1}, Y_{w,1},\cdots, Y_{w,J}$
are all independent Poisson processes
\cite{anderson2011continuous}.
The quantities 
\\ $R_{f_0}(\eta_0(t)), \cdots, R_{f,J-1}(\eta_{J-1}(t))$ are the fracture rates for fracturing a floe
of size category $0, \cdots, J-1$,
and the quantities $R_{w_1}(\eta_1(t)), \cdots, R_{w,J}(\eta_{J}(t))$ are the welding rates for welding two floes
of size category $1, \cdots, J$.
For each type of fracture and welding event, there is a corresponding term on the right-hand side
in the equation above. The times of each fracture or welding event are randomly occurring
according to an independent Poisson process.

Note that it is assumed above that a fracture event will
fracture one floe into two floes of equal size,
and a welding event will weld two equal-sized floes into
one floe. 
I.e., a welding event will weld two floes
from the same size category, rather than two floes from
different size categories.
In addition, it is assumed that there is a largest size category ($j=0$)
and a smallest size category ($j=J$),
and the largest floes do not weld together,
and the smallest floes do not fracture.
Other specifications could potentially be used instead,
although it is not clear if other specifications would lead to
an exact solution in power-law form.

Also note that,
for each particular size category $j$, the evolution equation has the same form:
\begin{align}
    \eta(j,t)
    &=
    \eta(j,0)
    -Y_{f,j}(\int_0^t R_{f,j}(\eta_j(t')) \, dt')
    +2Y_{f,j-1}(\int_0^t R_{f,j-1}(\eta_{j-1}(t')) \, dt')
    \nonumber \\
    &\qquad \qquad
    -2Y_{w,j}(\int_0^t R_{w,j}(\eta_j(t')) \, dt')
    +Y_{w,j+1}(\int_0^t R_{w,j+1}(\eta_{j+1}(t')) \, dt'),
    \label{eqn:eta-evol-j}
\end{align}
for $j=1,2,\cdots,J-1$.
As a result, the right-hand side includes only four jump processes
for each particular size category $j$.

Now we would like to take the expected value of 
$\eta(j,t)$ to consider $\mathbb{E}[\eta(j,t)]$.
Its evolution can be found by taking the expected value of
\eqref{eqn:eta-evol-j},
and using the property of the Poisson process $Y$ that
\begin{equation}
    \mathbb{E}[Y(\int_0^t R(t')\,dt')]
    =\mathbb{E}[\int_0^t R(t')\,dt'].
\end{equation}
The result is
\begin{align}
    \mathbb{E}[\eta(j,t)]
    &=
    \mathbb{E}[\eta(j,0)]
    -\int_0^t\mathbb{E}[R_{f,j}(\eta_j(t'))] \, dt'
    +2\int_0^t\mathbb{E}[R_{f,j-1}(\eta_{j-1}(t'))] \, dt'
    \nonumber \\
    &\qquad \qquad
    -2\int_0^t\mathbb{E}[R_{w,j}(\eta_j(t'))] \, dt'
    +\int_0^t\mathbb{E}[R_{w,j+1}(\eta_{j+1}(t'))] \, dt'.
    \label{eqn:eta-evol-expect}
\end{align}
This equation is potentially complicated, due to the 
potential nonlinear dependence on $\eta$ in the expressions
such as $\mathbb{E}[R_{f,j}(\eta_j(t))]$.

The following simplification can potentially be appropriate
for the case of sea ice:
linear reaction rates, i.e., $R_{f,j}(\eta_j(t))=r_{f,j}\eta_j(t)$
and $R_{w,j}(\eta_j(t))=r_{w,j}\eta_j(t)$
with constants $r_{f,j}$ and $r_{w,j}$, a case which occurs, e.g., 
in first-order chemical reactions
\cite{anderson2011continuous}.
Linear reaction rates occur, e.g., in unary chemical reactions,
i.e., where the reaction rate depends on
the concentration of only a single reactant.
This is in contrast to
binary chemical reactions, where the reaction rate
depends on the concentrations of
two reactants. 
For fracture and welding of sea ice, each size category of sea ice
represents a distinct type of ``molecule'' or ``chemical'' in this analogy.
A fracture event can naturally be viewed as a unary reaction,
since the reactant is a single floe.

A welding event can be viewed as 
having a linear reaction rate
under certain assumptions.
In this scenario, the rate of occurrence of welding events
depends linearly on the 
number of floes in a size category.
An alternative scenario would be
a chemical reaction where two molecules must
collide to initiate a chemical reaction and in which case
the reaction rate has a quadratic dependence on the
number of molecules.
For the welding of floes, the floes do not behave like
the molecules of a gas or liquid that must undergo rare collisions
in order to initiate a chemical reaction. 
Instead, floes are commonly
in contact with other floes, so that every floe
is often experiencing the conditions for welding to occur
(as opposed to molecules of a gas, which must undergo
a rare collision to initiate a reaction,
and for which the reaction rate is quadratic).
Hence, as a physical interpretation for floe welding,
we hypothesize
that the welding rate depends linearly on the number of floes
in a size category, so that each individual floe
experiences the same constant welding rate,
which is mainly determined by environmental factors
(such as temperature). In addition to this justification
based on physical characteristics of sea ice, further support for this setup
is provided by the statistics of welding events in
the SubZero DEM simulations described in the main text.

Based on the rationale above for
linear reaction rates in the case of sea ice, we will use
\begin{equation}
    R_{f,j}(\eta_j(t))=r_{f,j}\eta_j(t),
    \quad
    R_{w,j}(\eta_j(t))=r_{w,j}\eta_j(t),
    \quad
    \mbox{for each} \; j,    
    \label{eqn:R-linear-def}
\end{equation}
where $r_{f,j}$ and $r_{w,j}$ are constants,
so that the reaction rates
$R_{f,j}(\eta_j(t))$
and $R_{f,j}(\eta_j(t))$
depend linearly on $\eta_j(t)$.
(Note that, in full generality, the linearity should break down
for small values of $\eta(j,t)$ equal to 0 or 1 or 2.
This is because it is necessary to have $\eta(j,t)\ge 1$
for a fracture event and $\eta(j,t)\ge 2$ for a welding event.
To include these cases in full generality, 
one can modify the specification of the transition probabilities
in \eqref{eqn:R-linear-def} to include the special scenarios
that arise when $\eta(j,t)$ is equal to 0 or 1 or 2.
Instead, here we will just assume that the values of $\eta(j,t)$
are always sufficiently large so that \eqref{eqn:R-linear-def}
is always true, in order to simplify the presentation.)

In this case of linear reaction rates, 
the evolution of $\mathbb{E}[\eta(j,t)]$ in \eqref{eqn:eta-evol-expect}
simplifies to
\begin{align}
    \mathbb{E}[\eta(j,t)]
    &=
    \mathbb{E}[\eta(j,0)]
    -r_{f,j}\int_0^t\mathbb{E}[\eta_j(t')] \, dt'
    +2r_{f,j-1}\int_0^t\mathbb{E}[\eta_{j-1}(t')] \, dt'
    \nonumber \\
    &\qquad \qquad
    -2r_{w,j}\int_0^t\mathbb{E}[\eta_j(t')] \, dt'
    +r_{w,j+1}\int_0^t\mathbb{E}[\eta_{j+1}(t')] \, dt',
\end{align}
which is equivalent to a linear, constant-coefficient system of
ordinary differential equations,
\begin{equation}
    \frac{df_j}{dt}
    =2r_{f,j-1}f_{j-1}-(r_{f,j}+2r_{w,j})f_j+r_{w,j+1}f_{j+1}\, ,
    \label{eqn:fj-evol}
\end{equation}
where we have defined a simplified notation
for the expected value of the floe-size distribution:
\begin{equation}
    f_j(t)=\mathbb{E}[\eta(j,t)].
\end{equation}

A further simplification is supported by simulations of a
discrete element model for sea ice, as shown in a figure in the main text:
the fracture and welding rates are independent of $j$, i.e.,
\begin{equation}
    r_{f,j}=r_f, \qquad r_{w,j}=r_w, \qquad \mbox{for all} \, j.
\end{equation}
Under this simplifying assumption, the evolution equation
in \eqref{eqn:fj-evol} becomes
\begin{equation}
    \frac{df_j}{dt}
    =2r_f f_{j-1}-(r_f+2r_w)f_j+r_w f_{j+1},
\end{equation}
which is the final result of the derivation: 
the evolution equation for $f_j(t)$ that was written in the main text.

\subsection{Derivation \#2}

The master equation or Kolmogorov forward equation
\cite{gardiner2021elements}
offers a second derivation of the evolution equation for $f_j(t)$ in
\eqref{eqn:fj-evol}.
The master equation describes the probability $p(n_0,n_1,n_2,\cdots,n_J,t)$,
which is the probability that, at time $t$,
there are $n_0$ floes of size category 0,
there are $n_1$ floes of size category 1,
etc.
To connect with the random-time-change representation in
(\ref{eqn:eta-vec-evol}) for the stochastic process for
the number of floes $\eta(j,t)$ of size category $j$,
note the definition of the probability of a state,
\begin{equation}
    p(n_0,n_1,\cdots,n_J,t)=\mathbb{P}\{ \, \eta(0,t)=n_0, \, \eta(1,t)=n_1, \, \cdots, \, \eta(J,t)=n_J \, \}.
\end{equation}
The master equation is the evolution equation for $p(n_0,n_1,\cdots,n_J,t)$,
and for the stochastic model of sea ice from the main text, it is given by
\begin{align}
    \partial_t p(n_0,\cdots,n_J,t)
    &=
    -\sum_{j=0}^{J-1} r_f n_j p(n_0,\cdots,n_J,t)
    \nonumber \\
    &\qquad -\sum_{j=1}^{J} r_w n_j p(n_0,\cdots,n_J,t)
    \nonumber \\
    &\qquad +\sum_{j=0}^{J-1} r_f (n_j+1) p(\cdots, n_{j-1}, n_j+1, n_{j+1}-2, n_{j+2}, \cdots, t)
    \nonumber \\
    &\qquad +\sum_{j=1}^{J} r_w (n_j+2) p(\cdots, n_{j-2}, n_{j-1}-1, n_j+2, n_{j+1}, \cdots, t).
    \label{eqn:p-evol}
\end{align}
From the master equation, one can then determine the evolution equation for
the expected value of the floe-size distribution, i.e., the mean 
\begin{equation}
    f_i(t)=\sum_{n_0=0}^\infty \cdots \sum_{n_J=0}^\infty n_i p(n_0,\cdots,n_J,t).
    \label{eqn:fi-def}
\end{equation}
Its evolution equation can be found by taking the mean of \eqref{eqn:p-evol},
i.e., by multiplying \eqref{eqn:p-evol} by $n_i$ 
and summing over all possible states:
\begin{align}
    \frac{df_i}{dt}
    &=
    -\sum_{n_0=0}^\infty \cdots \sum_{n_J=0}^\infty n_i\sum_{j=0}^{J-1} r_f n_j p(n_0,\cdots,n_J,t)
    \nonumber \\
    &\qquad -\sum_{n_0=0}^\infty \cdots \sum_{n_J=0}^\infty n_i\sum_{j=1}^{J} r_w n_j p(n_0,\cdots,n_J,t)
    \nonumber \\
    &\qquad +\sum_{n_0=0}^\infty \cdots \sum_{n_J=0}^\infty n_i\sum_{j=0}^{J-1} r_f (n_j+1) p(\cdots, n_{j-1}, n_j+1, n_{j+1}-2, n_{j+2}, \cdots, t)
    \nonumber \\
    &\qquad +\sum_{n_0=0}^\infty \cdots \sum_{n_J=0}^\infty n_i\sum_{j=1}^{J} r_w (n_j+2) p(\cdots, n_{j-2}, n_{j-1}-1, n_j+2, n_{j+1}, \cdots, t).
    \label{eqn:dfidt-with-sums}
\end{align}
To simplify further, use a change of variable for the 
dummy variables in the summation.
For example,
for the third term on the right-hand side of \eqref{eqn:dfidt-with-sums},
use
\begin{gather}
    n_j' = n_j+1, \quad n_{j+1}'=n_{j+1}-2,
    \nonumber \\
    \mbox{and}\quad n_i'=n_i+1 \quad\mbox{if}\quad i=j,
    \nonumber \\
    n_i'=n_i-2 \quad\mbox{if}\quad i=j+1,
    \nonumber \\
    \quad n_i'=n_i \quad\mbox{if}\quad i<j \quad\mbox{or}\quad i>j+1.   
    \label{eqn:change-var-1}
\end{gather}
Then, to rewrite the third term on the right-hand side
of \eqref{eqn:dfidt-with-sums}, the change of variables in \eqref{eqn:change-var-1} is implemented as
\begin{align}
    & n_i\sum_{j=0}^{J-1} r_f (n_j+1) p(\cdots, n_{j-1}, n_j+1, n_{j+1}-2, n_{j+2}, \cdots, t)
    \nonumber \\
    &\qquad = r_f\sum_{j=0}^{i-2} n_i' n_j' p(\cdots, n_{j-1}, n_j', n_{j+1}', n_{j+2}, \cdots, t)
    \nonumber \\
    &\qquad\qquad +r_f (n_i'+2) n_j' p(\cdots, n_{j-1}, n_j', n_{j+1}', n_{j+2}, \cdots, t)|_{j=i-1}
    \nonumber \\
    &\qquad\qquad +r_f (n_i'-1) n_j' p(\cdots, n_{j-1}, n_j', n_{j+1}', n_{j+2}, \cdots, t)|_{j=i}
    \nonumber \\
    &\qquad\qquad +r_f\sum_{j=i+1}^{J-1} n_i' n_j' p(\cdots, n_{j-1}, n_j', n_{j+1}', n_{j+2}, \cdots, t)
    \nonumber \\
    &\qquad =r_f\sum_{j=0}^{J-1} n_i' n_j' p(n_0,\cdots,n_{j-1},n_j',n_{j+1}',n_{j+2},\cdots,n_J,t)
    \nonumber \\
    &\qquad\qquad -r_f n_i' p(n_0,\cdots,n_{i-1},n_i',n_{i+1}',n_{i+2},\cdots,n_J,t)
    \nonumber \\
    &\qquad\qquad +2r_f n_{i-1}' p(n_0,\cdots,n_{i-2},n_{i-1}',n_i',n_{i+1},\cdots,n_J,t).
    \label{eqn:dfidt-change-var-1}
\end{align}
Note that the final form involves a sum of terms with quadratic product 
$n_i'n_j'$, plus two additional terms that are linear
and proportional to either $n_i'$ or $n_{i-1}'$.
Similarly,
for the fourth term on the right-hand side of \eqref{eqn:dfidt-with-sums},
use the change of variables
\begin{gather}
    n_{j-1}' = n_{j-1}-1, \; n_j'=n_j+2,
    \nonumber \\
    \mbox{and}\quad n_i'=n_i-1 \quad\mbox{if}\quad i=j-1,
    \nonumber \\
    n_i'=n_i+2 \quad\mbox{if}\quad i=j,
    \nonumber \\
    \quad n_i'=n_i \quad\mbox{if}\quad i<j-1 \quad\mbox{or}\quad i>j.   
    \label{eqn:change-var-2}
\end{gather}
With these changes of variables described in 
\eqref{eqn:change-var-1}--\eqref{eqn:change-var-2},
the evolution equation in \eqref{eqn:dfidt-with-sums} can be written as
\begin{align}
    \frac{df_i}{dt}
    &=
    -r_f \sum_{j=0}^{J-1}\sum_{n_0=0}^\infty \cdots \sum_{n_J=0}^\infty n_i n_j p(n_0,\cdots,n_J,t)
    \nonumber \\
    &\qquad -r_w\sum_{j=1}^{J}\sum_{n_0=0}^\infty \cdots \sum_{n_J=0}^\infty n_i n_j p(n_0,\cdots,n_J,t)
    \nonumber \\
    &\qquad +r_f\sum_{j=0}^{J-1}\sum_{n_0=0}^\infty \cdots \sum_{n_{j-1}=0}^\infty \sum_{n_j'=0}^\infty \sum_{n_{j+1}'=0}^\infty \sum_{n_{j+2}=0}^\infty \cdots \sum_{n_J=0}^\infty n_i' n_j' p(n_0,\cdots,n_{j-1},n_j',n_{j+1}',n_{j+2},\cdots,n_J,t)
    \nonumber \\
    &\qquad -r_f\sum_{n_0=0}^\infty \cdots \sum_{n_{i-1}=0}^\infty \sum_{n_i'=0}^\infty \sum_{n_{i+1}'=0}^\infty \sum_{n_{i+2}=0}^\infty \cdots \sum_{n_J=0}^\infty n_i' p(n_0,\cdots,n_{i-1},n_i',n_{i+1}',n_{i+2},\cdots,n_J,t)
    \nonumber \\
    &\qquad +2r_f\sum_{n_0=0}^\infty \cdots \sum_{n_{i-1}=0}^\infty \sum_{n_i'=0}^\infty \sum_{n_{i+1}'=0}^\infty \sum_{n_{i+2}=0}^\infty \cdots \sum_{n_J=0}^\infty n_{i-1}' p(n_0,\cdots,n_{i-2},n_{i-1}',n_i',n_{i+1},\cdots,n_J,t)
    \nonumber \\
    &\qquad +r_w\sum_{j=1}^{J}\sum_{n_0=0}^\infty \cdots \sum_{n_{j-2}=0}^\infty \sum_{n_{j-1}'=0}^\infty \sum_{n_j'=0}^\infty \sum_{n_{j+1}=0}^\infty \cdots \sum_{n_J=0}^\infty n_i' n_j' p(n_0,\cdots,n_{j-2},n_{j-1}',n_j',n_{j+1},\cdots,n_J,t)
    \nonumber \\
    &\qquad +r_w\sum_{n_0=0}^\infty \cdots \sum_{n_{i-2}=0}^\infty \sum_{n_{i-1}'=0}^\infty \sum_{n_i'=0}^\infty \sum_{n_{i+1}=0}^\infty \cdots \sum_{n_J=0}^\infty n_{i+1}' p(n_0,\cdots,n_{i-1},n_i',n_{i+1}',n_{i+2},\cdots,n_J,t),
    \nonumber \\
    &\qquad -2r_w\sum_{n_0=0}^\infty \cdots \sum_{n_{i-2}=0}^\infty \sum_{n_{i-1}'=0}^\infty \sum_{n_i'=0}^\infty \sum_{n_{i+1}=0}^\infty \cdots \sum_{n_J=0}^\infty n_i' p(n_0,\cdots,n_{i-2},n_{i-1}',n_i',n_{i+1},\cdots,n_J,t).
    \label{eqn:dfidt-with-sums-dummy}
\end{align}
Note that \eqref{eqn:dfidt-with-sums-dummy} has eight terms on its
right-hand side. Four of the terms look somewhat similar to the four terms
from \eqref{eqn:dfidt-with-sums}, and they now involve quadratic products
of $n_in_j$ or $n_i'n_j'$. These quadratic terms all cancel:
the two terms with $\sum_{j=0}^{J-1}$ exactly cancel each other,
and the two terms with $\sum_{j=1}^J$ exactly cancel each other.
After removing these four terms, which cancel each other, we are left with
\begin{align}
    \frac{df_i}{dt}
    &=
    -r_f\sum_{n_0=0}^\infty \cdots \sum_{n_{i-1}=0}^\infty \sum_{n_i'=0}^\infty \sum_{n_{i+1}'=0}^\infty \sum_{n_{i+2}=0}^\infty \cdots \sum_{n_J=0}^\infty n_i' p(n_0,\cdots,n_{i-1},n_i',n_{i+1}',n_{i+2},\cdots,n_J,t)
    \nonumber \\
    &\qquad +2r_f\sum_{n_0=0}^\infty \cdots \sum_{n_{i-1}=0}^\infty \sum_{n_i'=0}^\infty \sum_{n_{i+1}'=0}^\infty \sum_{n_{i+2}=0}^\infty \cdots \sum_{n_J=0}^\infty n_{i-1}' p(n_0,\cdots,n_{i-2},n_{i-1}',n_i',n_{i+1},\cdots,n_J,t)
    \nonumber \\
    &\qquad +r_w\sum_{n_0=0}^\infty \cdots \sum_{n_{i-2}=0}^\infty \sum_{n_{i-1}'=0}^\infty \sum_{n_i'=0}^\infty \sum_{n_{i+1}=0}^\infty \cdots \sum_{n_J=0}^\infty n_{i+1}' p(n_0,\cdots,n_{i-1},n_i',n_{i+1}',n_{i+2},\cdots,n_J,t),
    \nonumber \\
    &\qquad -2r_w\sum_{n_0=0}^\infty \cdots \sum_{n_{i-2}=0}^\infty \sum_{n_{i-1}'=0}^\infty \sum_{n_i'=0}^\infty \sum_{n_{i+1}=0}^\infty \cdots \sum_{n_J=0}^\infty n_i' p(n_0,\cdots,n_{i-2},n_{i-1}',n_i',n_{i+1},\cdots,n_J,t).
    \label{eqn:dfidt-with-sums-dummy-four}
\end{align}
The remaining four terms on the right-hand side
are linear in $n_{i-1}'$, $n_i'$, or $n_{i+1}'$,
and they do not involve the sum
$\sum_{j=0}^{J-1}$ or $\sum_{j=1}^J$; they arose from the change of variables
for the dummy variables in the summations, as described in
\eqref{eqn:change-var-1}--\eqref{eqn:change-var-2}.
Each of the terms in \eqref{eqn:dfidt-with-sums-dummy-four}
involves a sum that can be identified
with either $f_{i-1}, f_i$, or $f_{i+1}$,
based on the definition in \eqref{eqn:fi-def}.
Hence \eqref{eqn:dfidt-with-sums-dummy-four} can be written as
\begin{equation}
    \frac{df_i}{dt}
    =2r_f f_{i-1}-(r_f+2r_w)f_i+r_w f_{i+1},
\end{equation}
which is the final result of the derivation: the evolution equation for $f_i(t)$ that was written in the main text.

\newpage
\section{Simulation setup: Stochastic sea ice floe model}

In this section, we describe the setup of numerical simulations of the
idealized stochastic model. These simulations provide numerical confirmation
and illustration of the theoretical results in the main text.

The stochastic simulation of a collection of floes is simulated, where the
state of the system is described by the quantity $\eta(j,t)$, as described
in the main text. Recall that $\eta(j,t)$ is the number of floes 
in size category $j$, with corresponding area
\begin{equation}
A_j = 4^{-j}.
\end{equation}
We choose to use area $A_j=4^{-j}$ instead of the $A_j=2^{-j}$
that was presented in the main derivation in the main text; 
this will serve to illustrate that the area
may be $A_j=c^{-j}$ for different $c$, and the theory in the main text
will hold with the same form with $\log_2$ changed to $\log_c$
with $c$ as the base of the logarithm,
as mentioned in the main text.
We consider 13 different areas of floes,
\begin{equation}
1, 4^{-1}, 4^{-2}, \cdots, 4^{-12},
\end{equation}
i.e.,
\begin{equation}
A_j = 4^{-j} \quad\mbox{for}\quad j=0,1,2,\cdots,12.
\end{equation}
The initial collection of floes is chosen to have
$\eta(j,t)|_{t=0}$ floes in size category $j$ as
\begin{equation}
\eta(j,t)|_{t=0} = 20\times 4^j, \quad\mbox{for}\quad j=0,1,2,\cdots,12.
\label{eqn:eta-init}
\end{equation} 
This setup is chosen to have 13 different sizes in order to allow a
vast range of scales, from a large-floe area of 1 to a small-floe area of
$4^{-12}\approx 6\times 10^{-8}$.
Moreover, the initial number of floes can be found by adding the
numbers of floes in all size categories $j$ based on \eqref{eqn:eta-init},
and it is initially a very large number:
$20\sum_{j=0}^{12} 4^{j}=20 (1-4^{13})(1-4)^{-1}\approx 4.47\times 10^{8}$.
With such a large range of scales
the numerical simulations are computationally costly.

\begin{figure}
\centering
\includegraphics[width=0.49\textwidth,  clip, trim={0.5cm 6.6cm 1.5cm 6.8cm}]{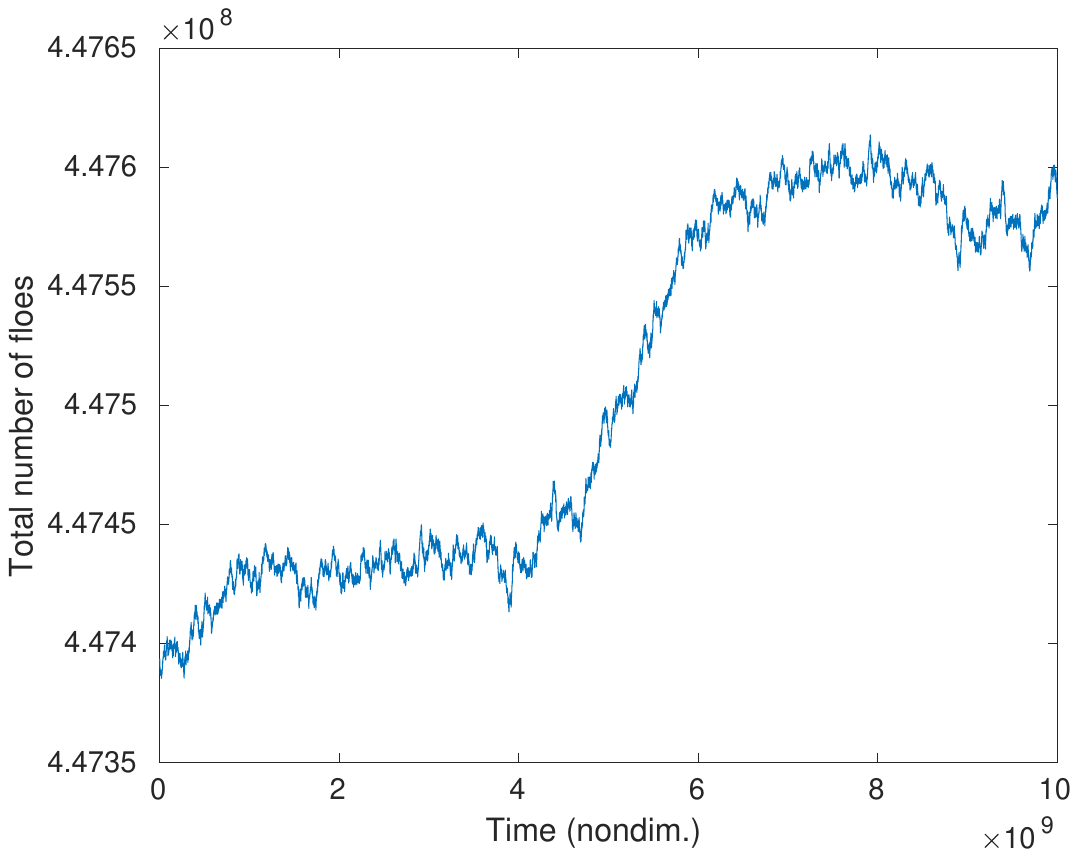}
\includegraphics[width=0.49\textwidth,  clip, trim={0.5cm 6.6cm 1.5cm 6.8cm}]{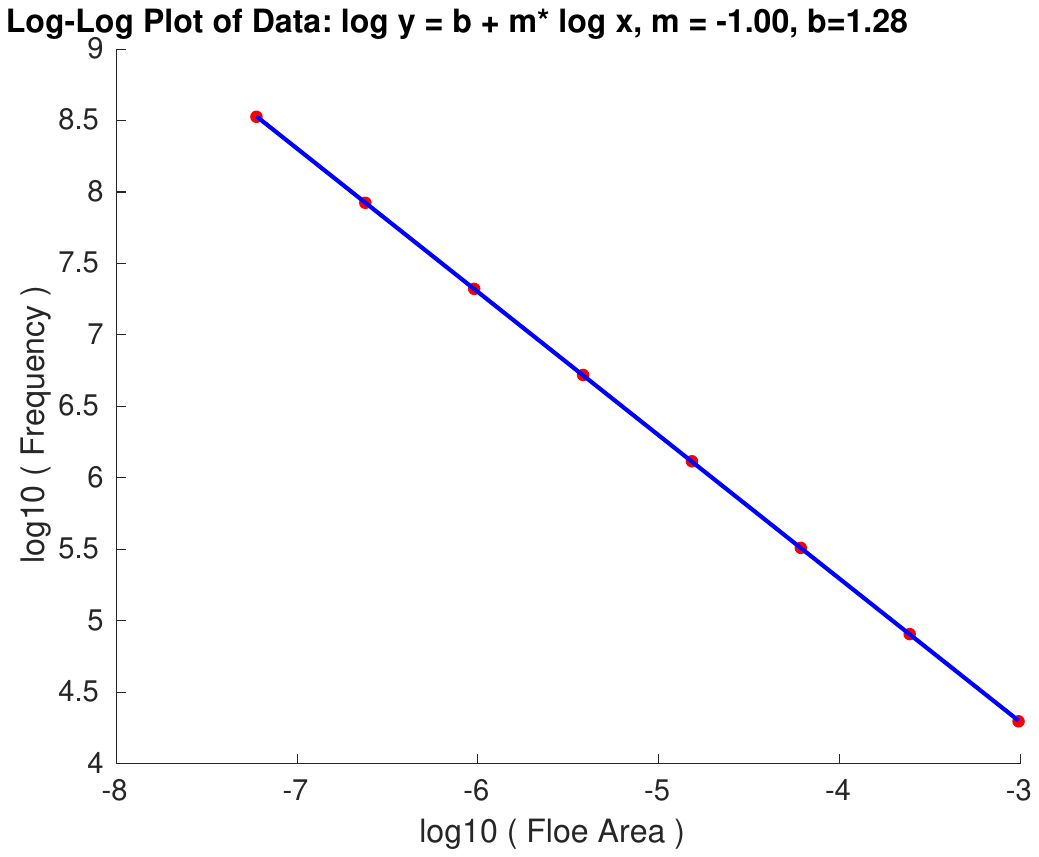}
\caption{Numerical simulation of the idealized stochastic model, for parameter values of $r_w = 0.05$ and $r_f = 0.2$. Left: Time series of the total number of floes in the system. Right: Log-log plot of the time-averaged FSD. Red symbols show the data from the simulation, and the blue line is a best-fit line. }
\label{fig:pf-point-2}
\end{figure}

The first simulation is initialized with the theoretically predicted
floe size distribution (FSD), and the time evolution illustrates the magnitude
of stochastic fluctuations. 
Recall that $\eta(j,t)$ has a stochastic evolution,
as described in the main text. 
At any instant of time, the 
FSD $\eta(j,t)$ is stochastic. As one indicator of the
stochastic fluctuations, Fig.~\ref{fig:pf-point-2} (left panel)
shows the fluctuations in the 
total number of floes in the system. The total number is
approximately 
$20\sum_{j=0}^{12} 4^j = 20\cdot (4^{13}-1)/(4-1)\approx 4.47\times 10^8$,
with fluctuations of magnitude of $10^5$ as time evolves over $10^{10}$
time steps.

\begin{figure}
\centering
\includegraphics[width=0.49\textwidth,  clip, trim={0.5cm 6.6cm 1.5cm 6.8cm}]{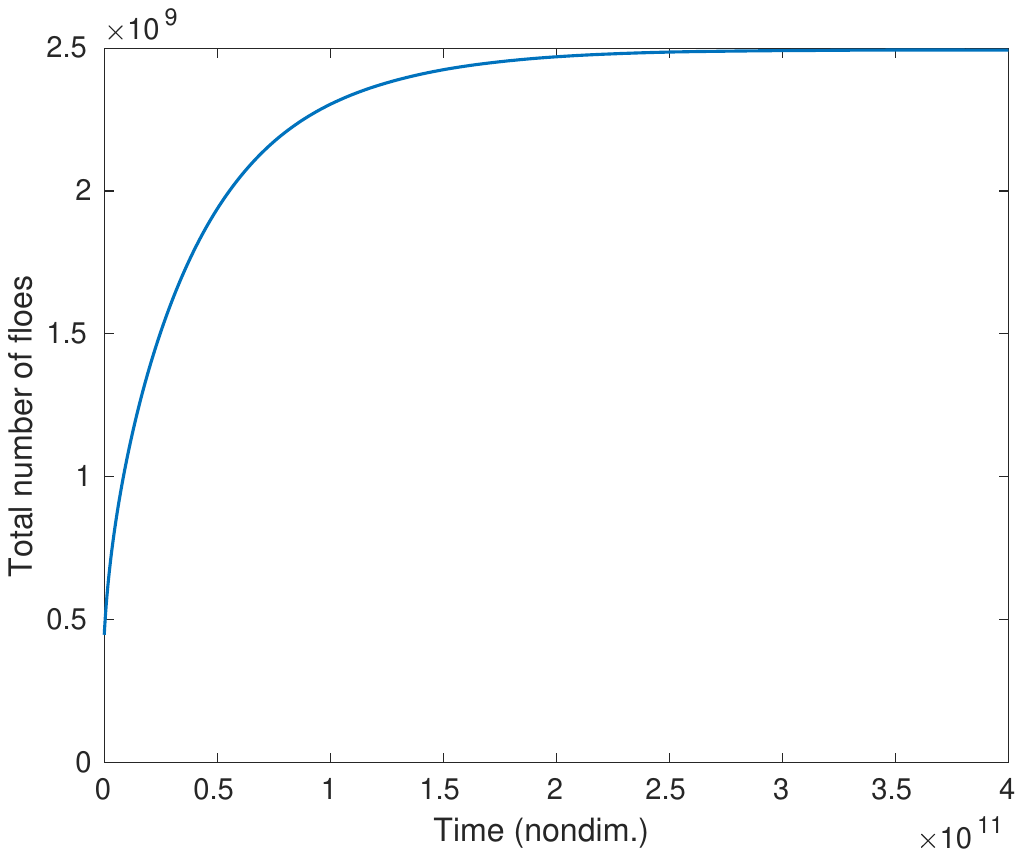}
\includegraphics[width=0.49\textwidth,  clip, trim={1.5cm 6.6cm 1.5cm 6.8cm}]{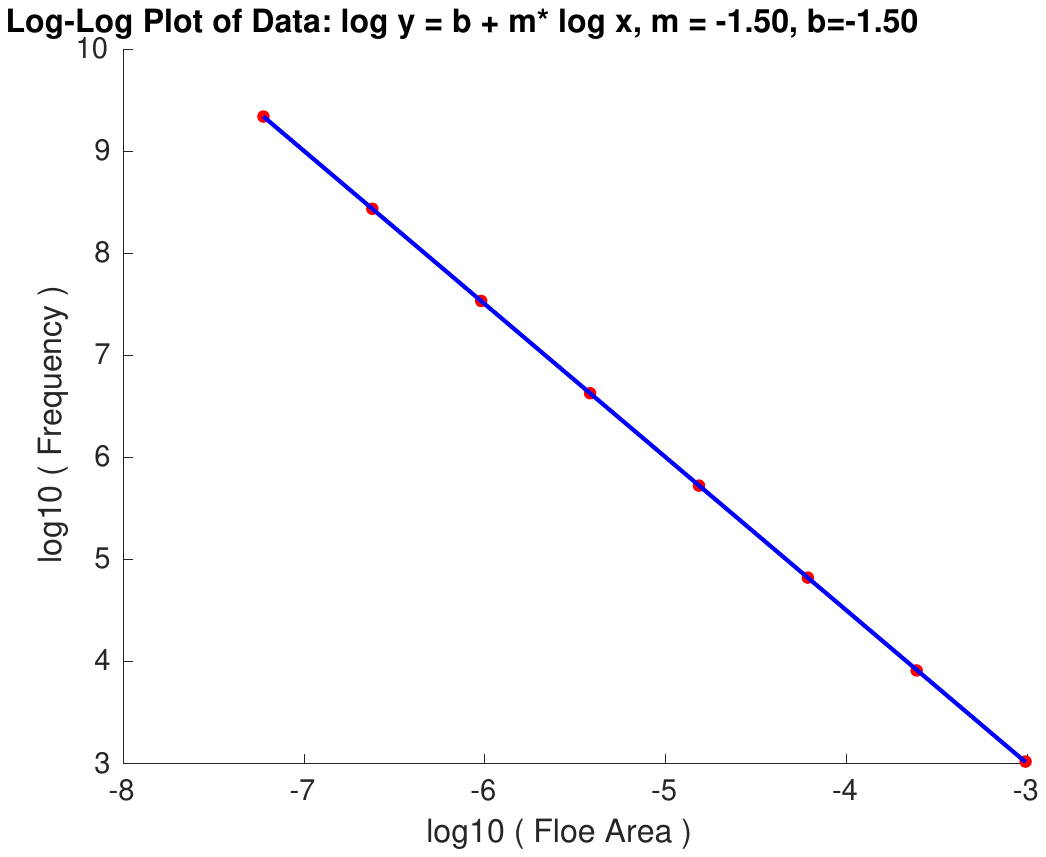}
\caption{Numerical simulation of the idealized stochastic model, for parameter values of $r_w = 0.05$ and $r_f = 0.4$. Left: Time series of the total number of floes in the system. Right: Log-log plot of the time-averaged FSD. Red symbols show the data from the simulation, and the blue line is a best-fit line.}
\label{fig:pf-point-4}
\end{figure}

\begin{figure}
\centering
\includegraphics[width=0.49\textwidth,  clip, trim={0.5cm 6.6cm 1.5cm 6.8cm}]{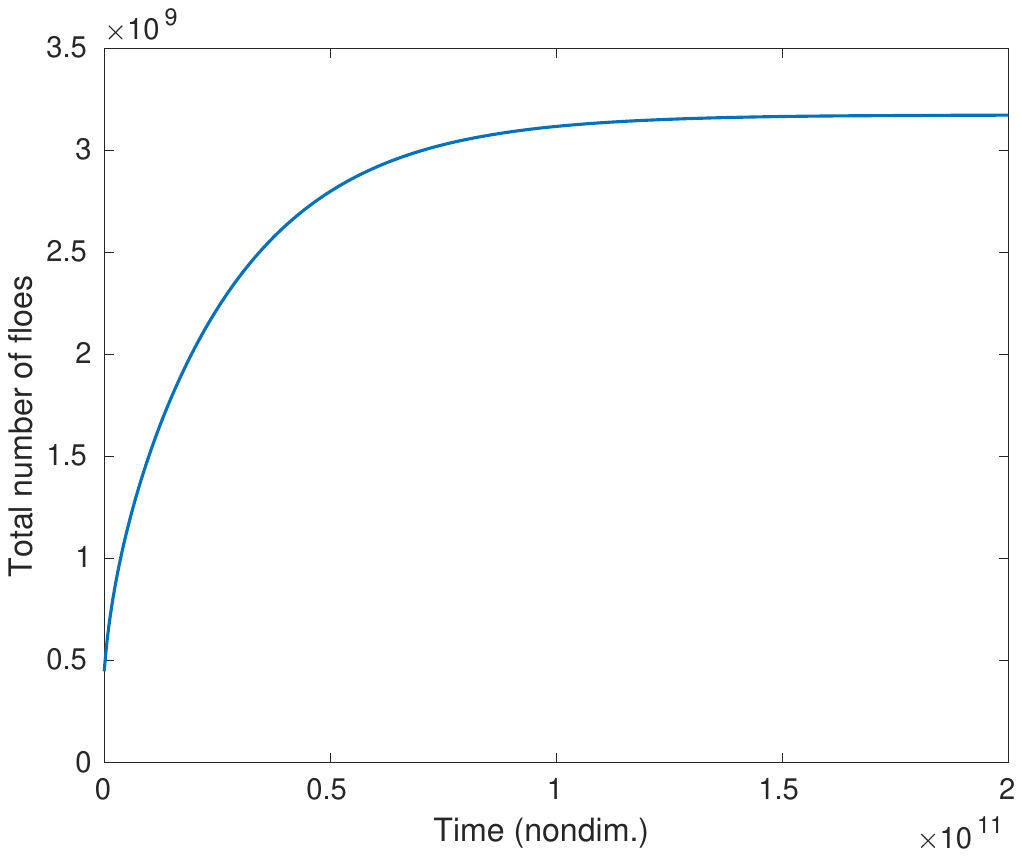}
\includegraphics[width=0.49\textwidth,  clip, trim={1.5cm 6.6cm 1.5cm 6.8cm}]{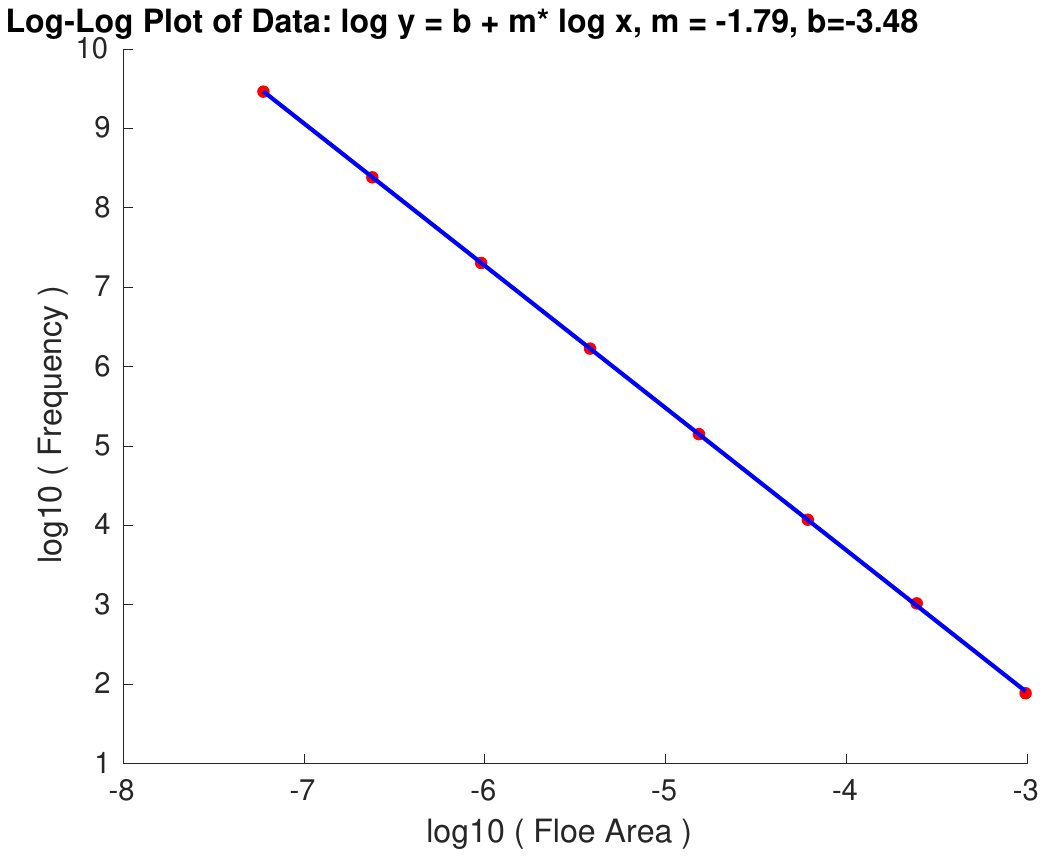}
\caption{Numerical simulation of the idealized stochastic model, for parameter values of $r_w = 0.05$ and $r_f = 0.6$. Left: Time series of the total number of floes in the system. Right: Log-log plot of the time-averaged FSD. Red symbols show the data from the simulation, and the blue line is a best-fit line.}
\label{fig:pf-point-6}
\end{figure}

\begin{figure}
\centering
\includegraphics[width=0.49\textwidth,  clip, trim={0.5cm 6.6cm 1.5cm 6.8cm}]{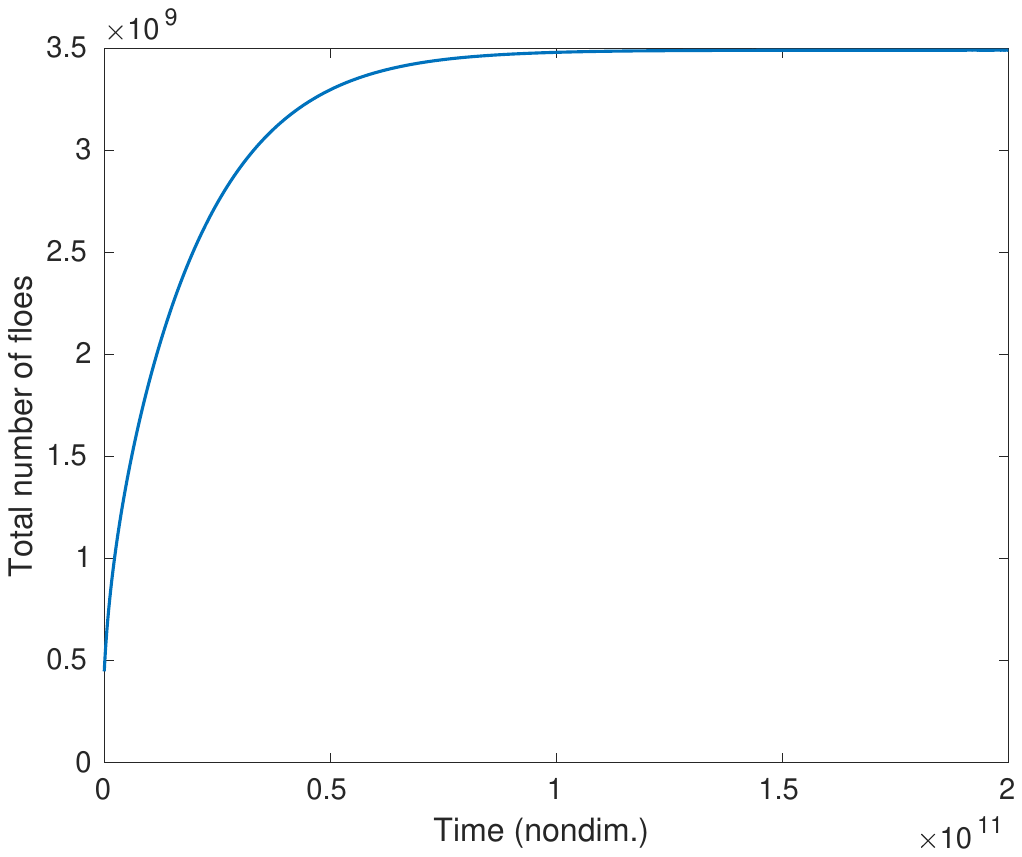}
\includegraphics[width=0.49\textwidth,  clip, trim={1.5cm 6.6cm 1.5cm 6.8cm}]{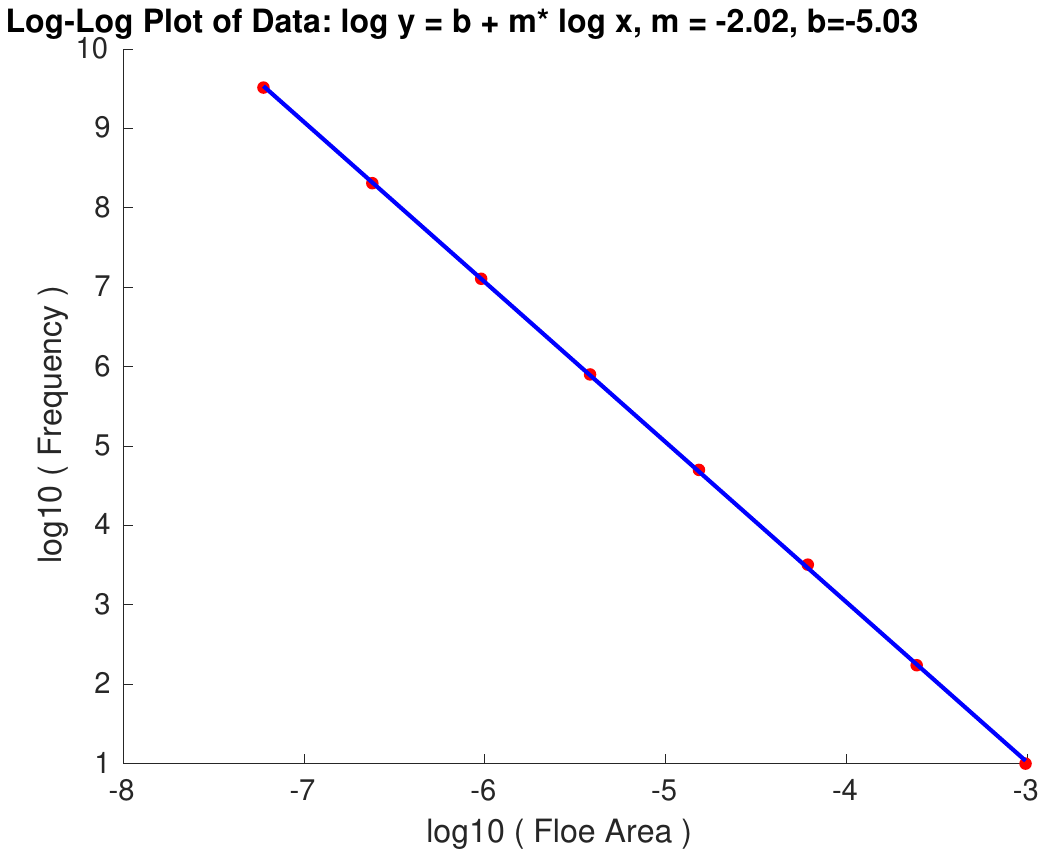}
\caption{Numerical simulation of the idealized stochastic model, for parameter values of $r_w = 0.05$ and $r_f = 0.8$. Left: Time series of the total number of floes in the system. Right: Log-log plot of the time-averaged FSD. Red symbols show the data from the simulation, and the blue line is a best-fit line.}
\label{fig:pf-point-8}
\end{figure}

In subsequent simulations 
in Figs.~\ref{fig:pf-point-4}--\ref{fig:pf-point-8} (left panels), 
in which different parameter values 
are used, the initial FSD is different from the equilibrium FSD, and
the adjustment to equilibrium takes place over about $10^{11}$
time steps.

The FSD in the stochastic simulations is shown 
in Figs.~\ref{fig:pf-point-2}--\ref{fig:pf-point-8} (right panels)
for different values of the fracture and welding rates, $r_f$ and $r_w$.
We keep the welding rate $r_w$ fixed at a low value of 0.05,
and we vary the fracture rate $r_f$ as 0.2, 0.4, 0.6, and 0.8. 
A different power law emerges in each case, and its exponent is in 
agreement with the theoretical predictions of 
$\alpha=\log_4 (r_f/r_w)=1.00, 1.50, 1.79,$ and 2.00,
respectively.
This is a numerical confirmation and illustration of the theoretical results
derived in the main text.

\clearpage
\section{Simulations with a fixed-rate version of the Subzero model}

In this section we describe the setup of the SubZero model in a fixed-rate
configuration. More specifically, the fracture and welding rates, per floe,
are taken to be fixed constants. This setup allows us to probe the
relationship between rates and FSD power laws in a DEM simulation.

The spatial domain is a square with a side length of $10^4$ and an area of $10^8$.
The units are dimensionless. To initialize the simulation, the ice begins as a
single sheet of ice of area $10^8$. Then, a succession of fracture events is
conducted to break the ice into 400 floes. The simulation then proceeds with
stochastic fracture and welding events. The fracture and welding rates,
per floe, have constant values of $r_f$ and $r_w$, respectively. The fracture
and welding rates of the whole collection of floes are $r_f\cdot (N_{all}-N_{small})$
and $r_w\cdot (N_{all}-N_{large})$, respectively, from adding up the rates
of the total number of floes, $N_{all}(t)$. Notice that small floes are assumed
not to fracture further, where $N_{small}$ is the number of floes with area
less than $10^4$, and large floes are assumed not to weld, where $N_{large}$
is the number of floes with area larger than $10^6$. Fracture events will break
one floe into two floes, and welding events will weld two floes into one floe.
Also note that the fracture and welding probabilities
of the collection depend on the (evolving) numbers of floes:
$N_{all}$, $N_{small}$, and $N_{large}$. The time stepping proceeds as 
follows. First, a random number is drawn to determine whether some event
occurs during the time step $\Delta t$, with probability
$[r_f\cdot (N_{all}-N_{small})+r_w\cdot (N_{all}-N_{large})]\Delta t$,
or whether no event occurs. If an event occurs, then a second random number
is drawn to determine whether a fracture event occurs or a welding event
occurs, with probabilities
$ r_f\cdot (N_{all}-N_{small}) [r_f\cdot (N_{all}-N_{small})+r_w\cdot (N_{all}-N_{large})]^{-1}$
and
$ r_w\cdot (N_{all}-N_{large}) [r_f\cdot (N_{all}-N_{small})+r_w\cdot (N_{all}-N_{large})]^{-1}$,
respectively. Lastly, a floe is selected at random, uniformly over all
$N_{all}-N_{small}$ eligible floes for a fracture or all
$N_{all}-N_{large}$ eligible floes for a welding event.
This procedure is repeated at each time step. The simulations are run forward
in time until a statistical equilibrium is reached.

\begin{figure}
\centering
\includegraphics[width=0.49\textwidth,  clip, trim={1.5cm 7.0cm 1.5cm 7.0cm}]{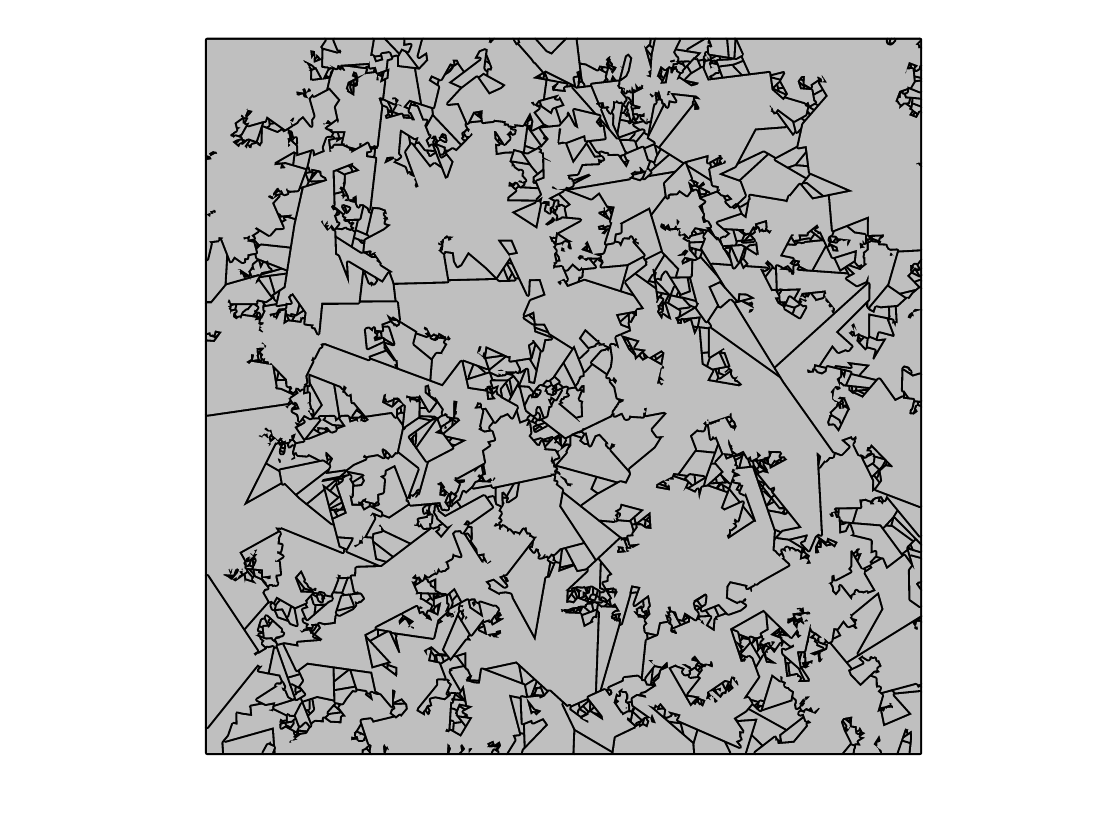}
\caption{Snapshot of floes from a simulation of the fixed-rate SubZero DEM in a square domain. The FSD is shown in Figure~\ref{fig:fixed-rate-subzero-power-laws}.}
\label{fig:fixed-rate-subzero-floes}
\end{figure}

\begin{figure} 
\centering 
\includegraphics[width=0.5\textwidth,  clip, trim={1.5cm 6.6cm 1.5cm 6.8cm}]{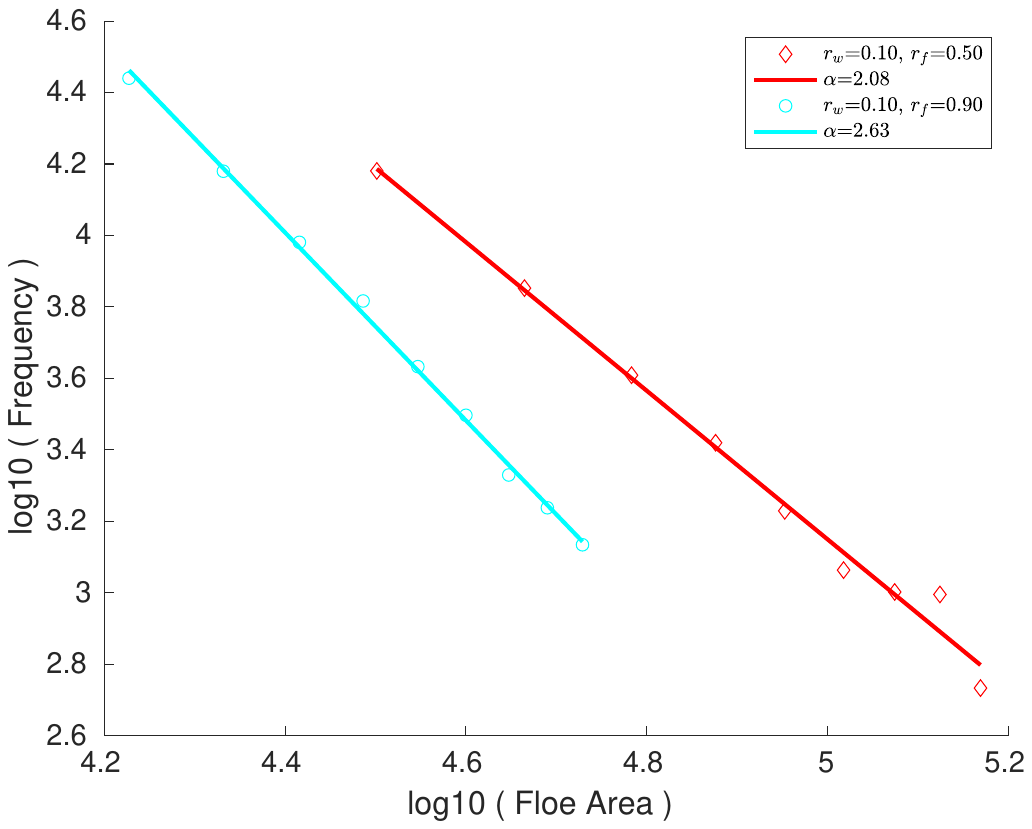} 
\caption{Log--log plot of the FSD from two simulations of the fixed-rate SubZero DEM. The power-law exponents $\alpha$ in these two simulations are
2.08 and 2.63, respectively, as indicated in the legend.
These values are roughly comparable to the theoretically predicted
values of $\log_2 (r_f/r_w)$ which are 2.32 and 3.17, respectively,
although there are deviations from the theoretically predicted values
due to various factors, including the additional complexity of
irregular floe shapes in the SubZero DEM model.}
\label{fig:fixed-rate-subzero-power-laws}
\end{figure}

The results of the simulations are shown in 
Figures~\ref{fig:fixed-rate-subzero-floes} and \ref{fig:fixed-rate-subzero-power-laws}.
A snapshot of the floes is shown in Figure~\ref{fig:fixed-rate-subzero-floes}, 
which illustrates the irregular shapes and the range of different floe sizes.
The FSD is illustrated in Figure~\ref{fig:fixed-rate-subzero-power-laws},
and it has the form of a power law.
Results are shown with 
the welding rate $r_w$ fixed at 0.1
and with the fracture rate $r_f$ fixed at either 0.5 or 0.9.
The power-law exponents $\alpha$ in these two simulations are
2.08 and 2.63, respectively.
These values are roughly comparable to the theoretically predicted
values of $\log_2 (r_f/r_w)$ which are 2.32 and 3.17, respectively,
although there are deviations from the theoretically predicted values
due to various factors, including the additional complexity of
irregular floe shapes in the SubZero DEM model.
In summary, these fixed-rate DEM simulations provide some further evidence 
that the general principles of the idealized stochastic model
may underlie the complex processes of the sea ice FSD in nature.

\clearpage

\end{document}